\newcommand{\CLASSINPUTbaselinestretch}{0.92}
\documentclass[conference]{IEEEtran}

\usepackage{cite}
\usepackage{amsmath}
\usepackage{stfloats}
\fnbelowfloat
\usepackage{url}

\ifCLASSINFOpdf

\else
    \usepackage{graphicx}
\fi

\begin{document}
%
\title{Short Paper: Design and Evaluation of Privacy-preserved 
Supply Chain System\\ based on Public Blockchain}


\author{
    \IEEEauthorblockN{Takio Uesugi\IEEEauthorrefmark{1},
        Yoshinobu Shijo\IEEEauthorrefmark{1}, 
        Masayuki Murata\IEEEauthorrefmark{1}
    }
    \IEEEauthorblockA{
        \IEEEauthorrefmark{1}Graduate School of Information Science and Technology, Osaka University \\
        1--5 Yamadaoka, Suita, Osaka, 565--0871 Japan\\
        Email: \{t-uesugi, y-shijo, murata\}@ist.osaka-u.ac.jp
    \vspace{-1pt}} 
}


\maketitle

\begin{abstract} 
Securing the traceability of products in the supply chain is an urgent issue.
Recently, supply chain systems that use public blockchain (PBC) have been proposed.
In these systems, PBC is used as a common database shared between 
supply chain parties to secure the integrity and reliability of distribution information such as ownership transfer records.
Thus, 
these systems secure a high level of traceability in the supply chain.
However, the distribution information, which can be private information, is made public since the information recorded in PBC can be read by anyone.
In this paper, we propose a method for preserving privacy while securing traceability in a supply chain system using PBC.
The proposed method preserves privacy by concealing the distribution information via encryption.
In addition, the proposed method ensures distribution among legitimate 
supply chain parties while concealing their blockchain address by using a zero-knowledge proof to prove their authenticity. 
We implement the proposed method on Ethereum smart contracts and evaluate cost performance based on transaction fees.
The results show that the fee per 
party is at most 2.6 USD.
\end{abstract}

%
\IEEEpeerreviewmaketitle

\section{Introduction}

\begin{figure*}[b]
    \centering
    \includegraphics[width=0.74\textwidth]{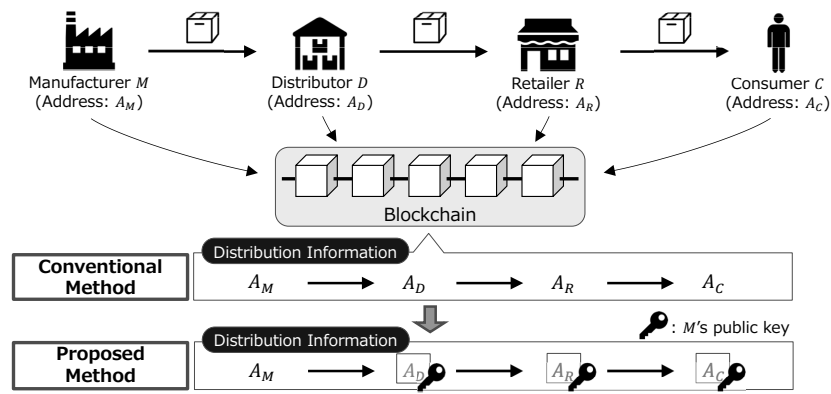}
    \caption{Overview of conventional and proposed methods.
    \label{fig:outline}}
\end{figure*}


Rapid globalization of supply chains has led to serious problems, particularly with respect to traceability.
The Organisation for Economic Co-operation and Development (OECD) reported that counterfeit products in international trade totaled 509 billion USD in 2016, up from 461 billion USD in 2013~\cite{OECD/EUIPO2019}.
Furthermore, due to the complexity of the supply chain, ingredients contaminated with \textit{Escherichia coli} could not be tracked, resulting in the 2015 \textit{E. coli} outbreak at Chipotle Mexican Grill~\cite{chipotle}.

To remedy these problems, supply chain systems based on blockchain have been proposed~\cite{Toyoda2017, TRU, HZL19}.
These systems store distribution information of products in a public blockchain (PBC).
Note that the distribution information in these systems are the records of ownership transfers.
Smart contracts prevent the registration of fraudulent distribution information by setting appropriate conditions.
Once information has been stored in the blockchain, no one can alter the information due to its immutability.
Therefore, these systems secure a high level of traceability in the supply chain.
Note that owners are managed by using their blockchain addresses.
In general, blockchain addresses are not tied to real-world entities.
Thus, a mechanism similar to public key certificates is used to link blockchain addresses to information about a company or individual.


However, there are privacy issues with these systems.
Distribution information is part of the competitive advantage of companies, and may include relationships between individuals in secondary markets.
Therefore, distribution information is private information that needs to be protected.
However, information recorded in a PBC can be freely viewed by anyone.
As a result, the privacy of distribution information is not protected.


In this paper, we propose a method for preserving privacy while securing traceability by the manufacturer of a product in supply chain systems that use PBC.
The main reason why conventional methods cannot preserve privacy is that they store the blockchain addresses directly in the blockchain.
Therefore, the proposed method preserves privacy by encrypting blockchain addresses by using the manufacturer's public key and storing the encrypted addresses in the blockchain, as shown in \figurename~\ref{fig:outline}.
The manufacturer can then track the product by decrypting the addresses using its own private key.
To eliminate 
distribution by 
illegitimate 
parties, the 
supply chain parties such as \textit{M, D, R} and \textit{C} in \figurename~\ref{fig:outline} 
must indicate to the system that they are legitimate 
parties when they ship and receive the product.
Accordingly, the proposed method uses a zero-knowledge proof to allow 
supply chain parties to prove their authenticity 
while hiding 
their blockchain addresses.

To evaluate 
the proposed method, 
we implement it on the Ethereum platform.
We assume 
distributions of a product, or transfers of ownership, starting with the manufacturer 
in the evaluation.
The results show that the proposed method can preserve privacy while securing traceability by the manufacturer.
We also evaluate the transaction fees for distribution and find that the fee per 
party is at most 2.6 USD.

The rest of this paper is structured as follows.
Section \ref{sec:related_work} discusses related work.
Section \ref{sec:proposed_method} introduces the proposed method, which is verified in Section \ref{sec:verification} in terms of privacy and traceability.
Section \ref{sec:evaluation} describes the environment for evaluating the proposed method and presents the evaluation results. 
We also present the results of evaluating transaction fees.
Lastly, our conclusions and future work are presented in Section \ref{sec:conclusions}.

\section{Related Work}
\label{sec:related_work}
Several blockchain-based systems have been proposed for improving the traceability of products in supply chains.
POMS~\cite{Toyoda2017} is a system for managing product ownership using the blockchain to prevent distribution of counterfeits in the post-supply chain.
Kim et al.~\cite{TRU} proposed a method for tracking products from the materials stage by repeated consumption and production of traceable resource units (TRUs).
Huang et al.~\cite{HZL19} proposes a method that can be applied to the food supply chain, which features high-frequency distribution, through the use of off-chain technology.
However, the protecting the privacy of distribution information has not been considered in any methods.

\section{Proposed Method}
\label{sec:proposed_method}

We propose a method for preserving the privacy of distribution information 
by extending POMS~\cite{Toyoda2017}.
The proposed method uses the public key of the product manufacturer to encrypt blockchain addresses.
Storing these encrypted addresses in the blockchain conceals the blockchain addresses and preserves privacy.
A manufacturer can track its products by using its own private key to decrypt and get the list of the blockchain addresses.
In addition, each of the 
supply chain parties proves, based on a zero-knowledge proof, that it knows a secret token possessed by only legitimate 
supply chain parties.
Thus, the proposed method ensures distribution among legitimate 
supply chain parties while concealing their blockchain addresses.

The proposed method consists of \textit{ManufacturerManagerContract (MMC)} for managing manufacturer information, \textit{ProductsManagerContract (PMC)} for managing product distributions, and \textit{VerifierContract (VC)} for verifying a proof based on the zero-knowledge proof.
In the following, we describe how to use these three contracts to prepare for distribution, register a product, manage distribution, and track the product.

\subsection{Preparation for distribution}
To prepare for distribution, the manufacturer information is registered in \textit{MMC}.
The required information is a pair consisting of the manufacturer's blockchain address and public key.
Other information such as the name and phone number of the manufacturer 
can also be registered if necessary.
In addition, \textit{MMC} associates the manufacturer's blockchain address with its products.

These 
registration processes 
can be executed only by a designated administrator.
For example, this could be performed by GS1~\cite{GS1}, an NPO that develops and maintains global standards for supply chains.


\subsection{Products registration}
Only a manufacturer registered with \textit{MMC} can register its products with \textit{PMC} as the first owner of the product.
Distribution of a product is initiated by a manufacturer's registration of the product information with \textit{PMC} after confirmation that the manufacturer is registered with \textit{MMC} and is associated with the product.
Below, \textit{PMC} is used to manage ownership.

\textit{PMC} records the raw blockchain address, not the encrypted address, for the manufacturer only.
This is because the manufacturer would like to prove that it has manufactured the product.
Anyone can freely identify the manufacturer of the product by viewing the first owner recorded in \textit{PMC}.

\subsection{Distribution management}
\begin{figure*}[t]
    \centering
    \includegraphics[width=0.8\textwidth]{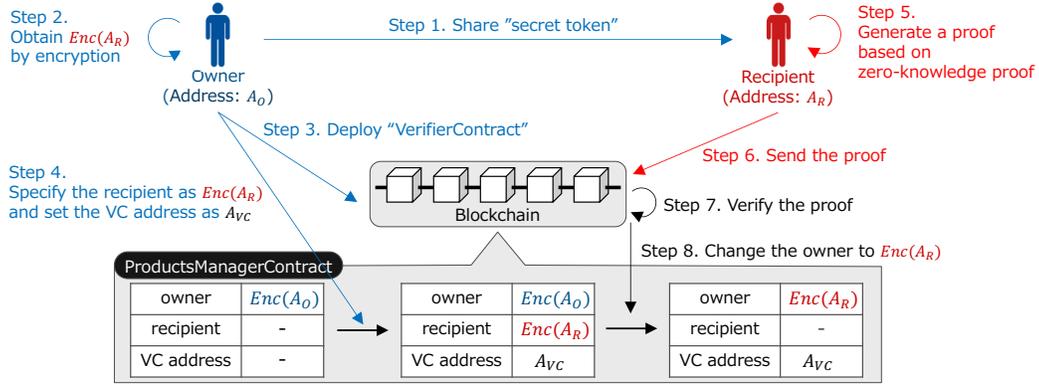}
    \caption{Flow of distribution management in the proposed method.
    \label{fig:flow-ProposedMethod}}
\end{figure*}

\figurename~\ref{fig:flow-ProposedMethod} illustrates the flow of distribution management.
Distribution management consists of the following eight steps: Steps~1 to~4 are the shipping process and Steps~5 to~8 are the receiving process.
\begin{enumerate}
    \item The owner shares a secret token with the recipient by a secure method.
    \item The owner encrypts the recipient's address $A_R$ using the secret token and the manufacturer's public key to obtain $\textit{Enc}(A_R)$.
    \item The owner deploys \textit{VC} on the blockchain.
    \item The owner records the recipient's encrypted address $\textit{Enc}(A_R)$ and the contract address of \textit{VC} obtained from Step~3 in \textit{PMC}.
    \item The recipient generates a proof that it knows the secret token shared in Step~1 based on a zero-knowledge proof.
    \item The recipient sends the proof to \textit{PMC}.
    \item \textit{PMC} calls \textit{VC} and verifies that the proof sent is valid.
    \item The owner is changed to $\textit{Enc}(A_R)$.
\end{enumerate}
\textit{PMC} provides the necessary functions to perform Steps 4, 6, 7, and 8.
The important steps are explained below.

First, we explain the encryption of the address in Step~2.
We use the elliptic curve elgamal encryption, as can be seen in (\ref{eq:ecelgamal}), to encrypt the address.
\begin{IEEEeqnarray}{rCl}
\textit{Enc}(A_R) = (kG, A_R + kQ) \label{eq:ecelgamal}
\end{IEEEeqnarray}
In the proposed method, $k$ is the secret token, $G$ is the generator of the elliptic curve, $A_R$ is the recipient's address, and $Q$ is the manufacturer's public key.
Note that $A_R$ must be a point on the elliptic curve for encryption. 
Therefore, we use a value converted from $A_R$ to a point on the elliptic curve. 

Second, we describe the zero-knowledge proof in Step~5.
To calculate the recipient's encrypted address $\textit{Enc}(A_R)$, the owner and the recipient must know all of $k, G, A_R$ and $Q$ in (\ref{eq:ecelgamal}).
Although $G, A_R$ and $Q$ are public information, $k$ is known only to the owner and the recipient.
That is, only the legitimate owner and recipient who know $k$ can correctly calculate $\textit{Enc}(A_R)$.
Therefore, the recipient can prove that it is legitimate by proving that it can calculate $\textit{Enc}(A_R)$.
We use a zero-knowledge proof for this purpose.
Zero-knowledge proof allows the recipient to prove that it can calculate $\textit{Enc}(A_R)$ without revealing $k$ and $A_R$.
Although there are various implementations of zero-knowledge proofs, we utilize zk-SNARKs which is known to be compatible with blockchain due to its non-interactivity and small proof size~\cite{zk-SNARKsInBC}. 
This proof is verified in Step~7.
\textit{PMC} calls \textit{VC} with the recipient's encrypted address $\textit{Enc}(A_R)$, the manufacturer's public key $Q$, and the proof generated in Step~5 as arguments to verify this proof.

In practice, even in Step~4, the owner proves that it is legitimate before recording the recipient's encrypted address in \textit{PMC}.
This is to prevent shipping by anyone other than the legitimate owner.
In the same way that the legitimate recipient is proved, the owner can prove that it is legitimate by proving that it can calculate $\textit{Enc}(A_O)$.

\subsection{Product tracking}
The blockchain records the owners'  encrypted  addresses as distribution information.
These are encrypted using the manufacturer's public key.
Therefore, the manufacturer can track the product by decrypting it using its own private key and arranging the decrypted addresses in chronological order.

\section{Verification}
\label{sec:verification}
We verify traceability and privacy of the proposed method by considering fraudulent activities by attackers.

\subsection{Traceability}
There are three possible attack vectors for inhibiting traceability.
The first is to interfere with decryption of the owner's encrypted address using the manufacturer's private key.
This can be performed by the attacker recording an encrypted statement on the blockchain using a public key other than the manufacturer's.
However, the proof verification in Step~7 is performed by directly referring to the manufacturer's public key recorded in the blockchain.
Therefore, the proof verification in Step~7 always fails if a statement encrypted with a public key other than the manufacturer's is recorded.
That is, distribution by an attacker also fails.
For this reason, it is impossible for this attack to succeed.

The second attack vector is for a third party not involved in distribution to carry out 
distribution by impersonating the owner or recipient.
This could happen if the third party is able to generate a valid proof.
However, those who do not know the information required for the proof cannot generate a valid proof because of the soundness of zero-knowledge proof.
Therefore, it is impossible for this attack to succeed.

The third attack vector is collusion between the owner and the recipient that makes it difficult to identify the owner.
This can be performed by the attacker using a random or real address other than his/her own.
The owner and recipient share an arbitrary address in addition to a secret token.
By using them to generate a proof, the owner information of \textit{PMC} is updated correctly. 
Therefore, it is possible for this attack to succeed.
We intend to deal with this attack in future work.

\subsection{Privacy}
An attacker can compromise privacy by retrieving the address from the owner's encrypted address or the proof recorded in \textit{PMC}.
First, we consider the encrypted address.
Cryptographic security depends on the key length and the cryptographic algorithm.
For example, when using 254-bit elliptic curve cryptography,  
it is extremely difficult for a party who does not know the private key to decrypt the encrypted address in practical time.
Thus, we can ensure the security of the encrypted address in the proposed method by using the 254-bit elliptic curve proposed in~\cite{bjj}.
Second, we consider the proof used in the proposed method.
We use a zero-knowledge proof in the proposed method that is known to satisfy zero-knowledge-ness.
In other words, it is not possible to recover information such as the address and the secret token from the proof.
As a result, the attacker cannot compromise privacy.
That is, the attacker cannot retrieve the owner's blockchain address.

\section{Evaluation}
\label{sec:evaluation}
To evaluate the proposed method, we implement it on the Ethereum platform. 
We also measure the transaction fees and discuss use cases based on the evaluation results.

\subsection{Environment setup}
We assume 
distributions of a 
product 
starting with the manufacturer.
The manufacturer executes product registration and shipping operations.
The other 
parties execute shipping and receiving operations.

We use Solidity version 0.5.11~\cite{Solidity} to write the smart contracts and the JavaScript Virtual Machine environment provided by Remix~\cite{Remix} to evaluate the 
proposed method.
We use ZoKrates~\cite{ZoKrates}, a toolbox of zk-SNARKs, for implementation of the zero-knowledge proof.

\begin{figure}[t]
    \centering
    \includegraphics[width=0.43\textwidth]{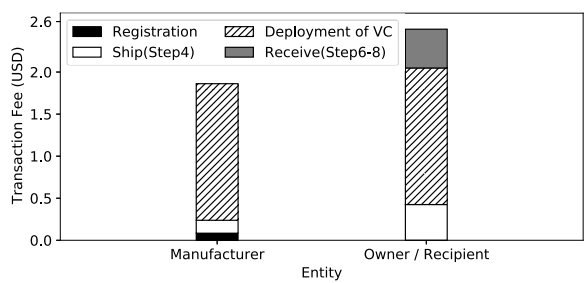}
    \caption{Transaction fee per party for a certain product.\label{fig:evaluation}}
\end{figure}

\subsection{Result and Discussion}
We found that the owner addresses could not be retrieved from the information recorded in 
\textit{PMC} and \textit{VC} after 
distributions of a 
product.
We also confirmed that the manufacturer could identify the product owner by decrypting the owner's encrypted addresses recorded in \textit{PMC} using its own private key.
The proposed method can preserve privacy while 
securing traceability by the manufacturer.

We measured the transaction fees for this distribution
and converted them into USD by multiplying the gas price by the gas value output by Remix.
At time of evaluation on March 17, 2020 at 11:00 a.m. (JST), the gas price was $1.1622 \times 10^{-6}$ USD per gas.
The maximum value of the transaction fee per party in this distribution is shown in \figurename~\ref{fig:evaluation}.
The manufacturer and the other 
parties have different 
transaction fees because they execute different processes.
In addition, the shipping process has different implementations of the functions executed by the manufacturer and the other 
parties, depending on whether the verification step is performed using the zero-knowledge proof.
Therefore, the transaction fees of the shipping process differ between the manufacturer and the other 
parties.

We found that the total transaction fees required for one party is at most 2.6 USD. 
Although there is room to reduce the fees by optimizing the implementation, there are still use cases that can be applied to high-priced products such as cars and home appliances even in the current implementation.
Such products are subject to recall if the product has a problem or defect.
The challenges of recalls include increasing consumer awareness of recalls and increasing recall response rates~\cite{OECD2018}.
In the case of distribution by the proposed method, only the manufacturer can track the product.
Therefore, these issues can be solved by tracking the product that is subject to recall and recalling that product through immediate notification of the owner.
The owner pays at most 2.6 USD to be able to promptly implement measures such as repair or replacement of the product. In this case, the transaction fee may be regarded as the fee for getting this kind of warranty and we believe that 2.6 USD is not an expensive charge for a warranty.

\section{Conclusions and Future Work}
\label{sec:conclusions}
In this paper, we proposed a method that can preserve privacy while securing traceability by the product manufacturer in a supply chain system using PBC.
We implemented the proposed method on the Ethereum platform and found the transaction fee per party to be at most 2.6 USD.

There are two issues that remain to be addressed in the future.
The first issue is consideration of ways to reduce the transaction fees.
Most of the transaction fees in the proposed method arise from the deployment process of \textit{VC}.
Instead of the owner deploying \textit{VC} for each distribution, the manufacturer could deploy it in advance.
If 
supply chain parties use a pre-deployed \textit{VC}, we can expect a significant reduction in the fees because of having only a one-time \textit{VC} deployment fee.
If the transaction fees can be reduced, we can apply this system to cheaper products.
The second issue is to extend the method so that it can be applied to the assembly and disassembly of products.
The proposed method assumes only the distribution of a single product without modification, that is, distribution of a finished product.
Therefore, if it could be applied to the assembly and disassembly of products, it could be applied to the distribution of products other than finished products.





\bibliographystyle{IEEEtran}
\bibliography{IEEEabrv, main}

\clearpage




\end{document}